\documentclass[aps,showpacs,twocolumn,preprintnumbers,amsmath,amssymb]{revtex4}
\usepackage{graphicx}
\usepackage{bm}

\def\br{ \bm{r} }
\def\bk{ \bm{k} }
\def\bo{ \bm{0} }
\def\im{ \,\mathrm{Im}\, }
\def\re{ \,\mathrm{Re}\, }

\def\bgam{ \bm{\gamma} }

\begin{document}

\title{NMR relaxation rate in non-centrosymmetric superconductors}

\author{K.~V.~Samokhin}

\address{Department of Physics, Brock University,
St.Catharines, Ontario, Canada L2S 3A1}
\date{\today}

\begin{abstract}
The spin-lattice relaxation rate of nuclear magnetic resonance in
a clean superconductor without inversion center is calculated for
arbitrary  pairing symmetry and band structure, in the presence of
strong spin-orbit coupling.
\end{abstract}


\maketitle

Most superconducting materials have an inversion center in their
crystal lattices. Among a few exceptions are the heavy-fermion
compounds CePt$_3$Si \cite{exp-CePtSi} and UIr \cite{exp-UIr}. The
peculiar feature of superconductivity without inversion center is
that there are fewer pairing channels available than in a
centrosymmetric crystal, because the spin degeneracy of the bands
is lifted in the presence of strong spin-orbit coupling
\cite{SZB04,SC04}. Other properties of non-centrosymmetric
superconductors studied recently include paramagnetic limit and
Knight shift \cite{BGR76,Edel89,GR01,FAKS04,FAS04,Sam05,Min05},
helical superconducting phases \cite{Edel96,Sam04,KAS04}, and
magnetoelectric phenomena \cite{LNE85,Edel95,Yip02}. The
determination of the pairing symmetry however presents a
considerable challenge. Although there are strong indications that
the gap in CePt$_3$Si has lines of nodes, the decisive proof,
including the determination of the nodal locations, is still
lacking.

A powerful probe of the quasiparticle properties in
superconductors is the spin-lattice relaxation rate $T_1^{-1}$ of
nuclear magnetic resonance (NMR). The presence of the coherence,
or Hebel-Slichter, peak in $T_1^{-1}$ below the critical
temperature \cite{HS59} provided an early strong support for the
Bardeen-Cooper-Schrieffer theory of superconductivity
\cite{Tink96}. More recently, the power-law behavior of the
relaxation rate at low temperatures commonly observed in
heavy-fermion compounds, see e.g. Ref. \cite{Flouq05}, has been
used as an argument in favor of the existence of gapless
excitations in those materials. Indeed, if there are line (point)
nodes in the gap, then $T_1^{-1}\propto T^3$ $(T^5)$ as $T\to 0$
\cite{SU91}.

The measurements of the NMR relaxation rate in CePt$_3$Si
\cite{Yogi04} indicate the likely presence of line nodes. On the
other hand, a small Hebel-Slichter peak just below $T_c$ observed
in Ref. \cite{Yogi04} was not found in other experiments
\cite{Ueda04}. Theoretically, the standard analysis of the NMR
relaxation rate in superconductors, see e.g. Ref. \cite{Tink96},
is not directly applicable in the non-centrosymmetric case because
of a complicated spin structure of the non-degenerate bands. The
purpose of the present Rapid Communication is to calculate
$T_1^{-1}$ in a non-centrosymmetric superconductor with arbitrary
pairing symmetry. Our analysis does not rely on any specific model
for the band structure, the only assumption being that the bands
are well split due to strong spin-orbit coupling, which is the
case for CePt$_3$Si \cite{SZB04}. In these circumstances, it is
convenient to use the exact band representation of the order
parameter introduced in Ref. \cite{SZB04}, see also Ref.
\cite{SC04}, in which the possibility of superconducting states
with lines of nodes at high-symmetry locations appears naturally.
In the alternative approach developed in Ref. \cite{FAKS04}, the
spin-orbit coupling is introduced using the Rashba model, and the
order parameter becomes a mixture of spin-singlet and spin-triplet
components, which does not in general have the lines of nodes
required by symmetry.

We consider the NMR spin-lattice relaxation due to the interaction
between the nuclear spin magnetic moment $\gamma_n\bm{I}$
($\gamma_n$ is the nuclear gyromagnetic ratio) and the hyperfine
field $\bm{h}$ created at the nucleus by the conduction electrons.
The system Hamiltonian is $H=H_e+H_n+H_{int}$, where $H_e$
describes the electron subsystem, $H_n=-\gamma_n\bm{I}\bm{H}$ is
the Zeeman coupling of the nuclear spin with the external field
$\bm{H}$, and $H_{int}=-\gamma_n\bm{I}\bm{h}$ is the hyperfine
interaction. For $I=1/2$, we have two nuclear spin states $I_z=\pm
1/2$ with the energies $E_{I_z}=-\omega_0 I_z$, where
$\omega_0=\gamma_nH$ is the NMR frequency (the spin quantization
axis is parallel to $\bm{H}$). Using the Fermi Golden Rule, the
spin-lattice relaxation rate can be expressed \cite{Slichter90} in
terms of the retarded correlator of the transverse components of
the hyperfine fields:
\begin{equation}
\label{R}
    R\equiv\frac{1}{T_1T}=-\frac{\gamma_n^2}{2\pi}
    \lim\limits_{\omega_0\to 0}
    \frac{\im K_{+-}^R(\omega_0)}{\omega_0}
\end{equation}
where $K_{+-}^R$ is obtained by an analytic continuation of the
Fourier transform of the Matsubara correlator
$K_{+-}(\tau)=-\langle T_\tau
e^{H_e\tau}h_+e^{-H_e\tau}h_-\rangle$, with $h_\pm=h_x\pm ih_y$
(in our units $k_B=\hbar=1$).

In general, the hyperfine field $\bm{h}$ can be represented as a
sum of the Fermi contact, the orbital, and the spin-dipolar
contributions \cite{Slichter90}. Their relative importance depends
on the electronic structure and therefore varies for different
systems. Let us first assume that the Fermi contact interaction is
dominant, then $\bm{h}=\alpha\bm{S}(\bo)$, where
$\alpha=-(8\pi/3)\gamma_e$, $\gamma_e$ is the electron
gyromagnetic ratio, and
$\bm{S}(\br)=(1/2)\bm{\sigma}_{\sigma\sigma'}\psi_\sigma^\dagger(\br)\psi_{\sigma'}(\br)$
is the electron spin density ($\bm{\sigma}$ is a vector composed
of Pauli matrices, and $\sigma=\uparrow,\downarrow$ is the spin
projection). The relaxation rate (\ref{R}) is then proportional to
the retarded correlator of the electron spin densities at the
nuclear site, which can be calculated by writing the spin
operators $S_\pm=S_x\pm iS_y$ in the band representation.

In a non-centrosymmetric crystal with spin-orbit coupling, the
electron bands are non-degenerate almost everywhere, except along
some high-symmetry lines in the Brillouin zone. Using the Bloch
theorem, the spinor field operators can be written as
\begin{equation}
\label{psis}
    \psi_\sigma(\br)=\frac{1}{\sqrt{V}}\sum\limits_{i,\bk}
        \chi_{i,\bk}(\br,\sigma)e^{i\bk\br}c_{i,\bk},
\end{equation}
where $V$ is the system volume, $i$ is the band index, $c_{i,\bk}$
are the destruction operators of band electrons, and
$\chi_{i,\bk}(\br,\sigma)=\langle\br,\sigma|i,\bk\rangle$ are
periodic functions in the unit cell, satisfying the normalization
condition
$$
    \frac{1}{\Omega}\int\limits_{\mathrm{unit\ cell}}d^3\br
    \left[|\chi_{i,\bk}(\br,\uparrow)|^2+
    |\chi_{i,\bk}(\br,\downarrow)|^2\right]=1
$$
($\Omega$ is the volume of the unit cell). The band energies are
$\epsilon_i(\bk)=\epsilon_i(-\bk)$. We introduce the notations
$u_{i,\bk,\sigma}=\chi_{i,\bk}(\bo,\sigma)$.

There exist certain relations between the Bloch amplitudes, which
follow from the invariance of the system with respect to time
reversal. The time-reversal operation $K=(i\sigma_2)K_0$ ($K_0$ is
the complex conjugation) transforms the Bloch spinor corresponding
to $\bk$ in band $i$ into a spinor corresponding to $-\bk$ in the
same band: $K|i,\bk\rangle=t_i(\bk)|i,-\bk\rangle$, where
$t_i(\bk)$ is a non-trivial phase factor. From $K^2=-1$ one
immediately obtains $t_i(-\bk)=-t_i(\bk)$ \cite{SC04}. Then it is
straightforward to show that
\begin{equation}
\label{ab relations}
    u_{i,-\bk,\uparrow}=t_i^*(\bk)u^{*}_{i,\bk,\downarrow},\quad
    u_{i,-\bk,\downarrow}=-t_i^*(\bk)u^{*}_{i,\bk,\uparrow}.
\end{equation}
Since the phases of the Bloch functions can be rotated
independently at different $\bk$, there is some freedom in
choosing $u_{i,\bk,\sigma}$ and $t_i(\bk)$. For this reason, no
observable quantity can depend on $t_i(\bk)$.

As an example, let us consider the generalized Rashba model, in
which the free-electron Hamiltonian is written as
$H_0=\epsilon_0(\bk)+\bgam(\bk)\bm{\sigma}$, where
$\epsilon_0(\bk)$ is an even function of $\bk$. The effects of
spin-orbit coupling are described by a real pseudovector
$\bgam(\bk)=-\bgam(-\bk)$, whose momentum dependence is determined
by the point symmetry of the crystal (the case of the tetragonal
group $\mathbf{C}_{4v}$ relevant for CePt$_3$Si is discussed in
Refs. \cite{FAKS04,Sam04,Sam05}). The diagonalization of $H_0$
gives two Rashba bands with the energies
$\epsilon_\pm(\bk)=\epsilon_0(\bk)\pm|\bgam(\bk)|$ and
$u_{\pm,\bk,\uparrow}=(1/\sqrt{2})\sqrt{1\pm\gamma_z/|\bgam|}$,
$u_{\pm,\bk,\downarrow}=\pm(\gamma_x+i\gamma_y)/\sqrt{2|\bgam|(|\bgam|\pm\gamma_z)}$.
Applying the time reversal operation, one obtains $t_\pm(\bk)=\pm
e^{-i\phi(\bk)}$, where $\phi=\arg(\gamma_x,\gamma_y,0)$. It is
easy to check that the relations (\ref{ab relations}) are indeed
satisfied.

The Hamiltonian $H_e$ of the electron subsystem consists of the
free-electron part and the interaction responsible for the Cooper
pairing. The pairs are formed by electrons in the degenerate
time-reversed states $|i,\bk\rangle$ and $K|i,\bk\rangle$. The
large band splitting strongly suppresses the pairing of electrons
from different bands. An anisotropic multi-band generalization of
the mean-field BCS model reads
\begin{eqnarray}
\label{He}
    H_e&=&\sum\limits_{i,\bk}\xi_{i,\bk}c^\dagger_{i,\bk}c_{i,\bk}\nonumber\\
    &&+\frac{1}{2}\sum\limits_{i,\bk}\left[\Delta_{i,\bk}c^\dagger_{i,\bk}
    c^\dagger_{i,-\bk}+\Delta^*_{i,\bk}c_{i,-\bk}c_{i,\bk}\right],
\end{eqnarray}
where the chemical potential $\mu$ is included in the
free-electron part, i.e. $\xi_{i,\bk}=\epsilon_i(\bk)-\mu$. The
pairing amplitudes $\Delta_{i,\bk}=-\Delta_{i,-\bk}$ can be
written as $\Delta_{i,\bk}=t_i(\bk)\tilde\Delta_{i,\bk}$, where
the auxiliary gap functions
$\tilde\Delta_{i,\bk}=\tilde\Delta_{i,-\bk}$ all have the same
symmetry, determined by an even irreducible representation
$\Gamma$ of the crystalline point group. The last observation
allows us to write
$\tilde\Delta_{i,\bk}=\sum_{a=1}^{d_\Gamma}\eta^{(i)}_a\phi^{(i)}_a(\bk)$,
where $d_\Gamma$ is the dimensionality of $\Gamma$,
$\phi^{(i)}_a(\bk)$ are the even basis functions (which can have
different $\bk$-dependence in different bands while having the
same transformation properties and sharing the same
symmetry-imposed zeros), and $\eta^{(i)}_a$ are the order
parameter components satisfying the mean-field self-consistency
equations \cite{Book}.

In the band representation, the spin raising and lowering
operators take the following form:
\begin{equation}
\label{Spm}
    S_+(\bo)=\frac{1}{V}\sum\limits_{ij,\bk\bk'}u^{*}_{i,\bk,\uparrow}
    u_{j,\bk',\downarrow}c^\dagger_{i,\bk}c_{j,\bk'},
\end{equation}
and $S_-(\bo)=S^\dagger_+(\bo)$. Inserting these expressions in
the Matsubara spin correlator one arrives at a two-particle
Green's function, which can be decoupled in the mean-field
approximation and represented in terms of the normal and anomalous
Green's functions for the Hamiltonian (\ref{He}): $-\langle T_\tau
c_{i,\bk}(\tau)c^\dagger_{j,\bk'}(0)\rangle=\delta_{ij}\delta_{\bk,\bk'}G_i(\bk,\tau)$
and $\langle T_\tau
c_{i,\bk}(\tau)c_{j,\bk'}(0)\rangle=\delta_{ij}\delta_{\bk,-\bk'}F_i(\bk,\tau)$.
Using the relations (\ref{ab relations}), we arrive at the
following expression for the spin correlator:
\begin{eqnarray}
\label{Kpm Matsubara}
    K_{+-}(\nu_m)=\alpha^2T\sum\limits_n\frac{1}{V^2}\sum\limits_{ij,\bk\bk'}
    |u_{i,\bk,\downarrow}|^2|u_{j,\bk',\uparrow}|^2\nonumber\\
    \times\bigl[G_i(\bk,\omega_n+\nu_m)G_j(\bk',\omega_n)\nonumber\\
    +t_i(\bk)t_j^*(\bk')F^\dagger_i(\bk,\omega_n+\nu_m)F_j(\bk',\omega_n)\bigr],
\end{eqnarray}
where $\omega_n=(2n+1)\pi T$ and $\nu_m=2m\pi T$ are the fermionic
and bosonic Matsubara frequencies respectively, and
\begin{eqnarray*}
    &&G_i(\bk,\omega_n)=-\frac{i\omega_n+\xi_{i,\bk}}{\omega_n^2+\xi^2_{i,\bk}+|\Delta_{i,\bk}|^2}\\
    &&F_i(\bk,\omega_n)=\frac{\Delta_{i,\bk}}{\omega_n^2+\xi^2_{i,\bk}+|\Delta_{i,\bk}|^2},
\end{eqnarray*}
$F^\dagger_i(\bk,\omega_n)=F^*_i(\bk,\omega_n)$. One can see that
the phase factors $t_i(\bk)$ drop out of the spin correlator
(\ref{Kpm Matsubara}), as expected.

The next steps are standard: first, we perform the summation over
$n$, followed by the analytical continuation
$K^R_{+-}(\omega)=K_{+-}(\nu_m)|_{i\nu_m\to\omega+i0^+}$, then
take the thermodynamic limit $V\to\infty$ and neglect the
electron-hole asymmetry, which allows us to write the momentum
sums as the Fermi-surface integrals. The final result for the
relaxation rate (\ref{R}) is
\begin{eqnarray}
\label{T1T result}
    R=\alpha^2\gamma_n^2\int\limits_0^\infty d\omega
    \left(-\frac{\partial f}{\partial\omega}\right)\bigr\{N_\uparrow(\omega)
    N_\downarrow(\omega)\nonumber\\
    +\re[M^*_\uparrow(\omega)M_\downarrow(\omega)]\bigr\},
\end{eqnarray}
where $f(\omega)=(e^{\omega/T}+1)^{-1}$ is the Fermi function, and
\begin{equation}
\label{NM}
    \left.\begin{array}{c}
    \displaystyle N_\sigma(\omega)=\sum\limits_iN_{F,i}\left\langle|u_{i,\bk,\sigma}|^2
    \frac{\omega}{\sqrt{\omega^2-|\tilde\Delta_{i,\bk}|^2}}
    \right\rangle\biggr._i,\\ \\
    \displaystyle M_\sigma(\omega)=\sum\limits_iN_{F,i}\left\langle|u_{i,\bk,\sigma}|^2
    \frac{\tilde\Delta_{i,\bk}}{\sqrt{\omega^2-|\tilde\Delta_{i,\bk}|^2}}
    \right\rangle\biggr._i.
    \end{array}\right.
\end{equation}
Here the angular brackets denote the Fermi-surface average, and
$N_{F,i}=(1/8\pi^3)\int dS_F/|\bm{v}_{F,i}|$ is the density of
states at the Fermi level in the $i$th band. The angular
integration is restricted by the condition
$|\tilde\Delta_{i,\bk}|\leq\omega$.

We see that the intra-band contributions to the NMR relaxation
rate in a non-centrosymmetric superconductor are similar to those
in the centrosymmetric case, despite a nontrivial spin structure
of the single-electron bands. The only difference is that it is
the auxiliary gap functions $\tilde\Delta_{i,\bk}$ that enter Eqs.
(\ref{NM}). It is important however that, in addition to the
intra-band terms, there are inter-band interference contributions
to $R$, which are present even without any inter-band scattering
mechanisms due to electron-electron interactions or impurities
\cite{SM05}. The origin of these contributions can be traced back
to the local character of the hyperfine coupling $\bm{I}\bm{S}$,
which mixes together the electron states near the Fermi surface
from different bands.

The low-temperature properties of the system are controlled by the
Bogoliubov quasiparticles of energy
$E_{i,\bk}=\sqrt{\xi^2_{i,\bk}+|\tilde\Delta_{i,\bk}|^2}$. The
experimental data for CePt$_3$Si seem to point to the presence of
lines of nodes in $\tilde\Delta_{i,\bk}$. Symmetry-imposed gap
nodes exist only for the order parameters which transform
according to one of the non-unity representations of the point
group. For all such non-trivial gap symmetries, the Fermi-surface
averages of $\tilde\Delta_{i,\bk}$ vanish, and therefore
$M_\uparrow(\omega)=M_\downarrow(\omega)=0$. Then the relaxation
rate in the superconducting normalized to its normal-state value
can be written as
\begin{equation}
\label{T1T nonunity}
    \frac{R_s}{R_n}=2\int\limits_0^\infty d\omega
    \left(-\frac{\partial f}{\partial\omega}\right)\frac{N_\uparrow(\omega)
    N_\downarrow(\omega)}{N_{n,\uparrow}N_{n,\downarrow}},
\end{equation}
where
$N_{n,\sigma}=\sum_iN_{F,i}\langle|u_{i,\bk,\sigma}|^2\rangle_i$
are the local densities of states of the spin-up (spin-down)
electrons at the nuclear site $\br=\bo$ in the normal state. Note
that, although $|u_{i,\bk,\uparrow}|^2+|u_{i,\bk,\downarrow}|^2=1$
in the generalized Rashba model, this does not have to be the case
in general, so that $N_{n,\uparrow}+N_{n,\downarrow}\neq N_F$,
where $N_F=\sum_iN_{F,i}$ is the total density of states at the
Fermi level for both spin projections.

Since the order parameters induced by the inter-band pair
scattering have the same symmetry in all bands, the low-energy
behavior of all terms in $N_\sigma(\omega)$, see Eq. (\ref{NM}),
is characterized by the same power law. If there are line (point)
nodes in the gap, then $N_\sigma(\omega)\propto\omega$
$(\omega^2)$ at $\omega\to 0$ \cite{Book}, and $R\propto T^2$
$(T^4)$ at $T\to 0$ \cite{SU91,Hase96}. However, a complicated
multi-band structure with considerably different local densities
of states and gap magnitudes in different bands can obscure the
simple power-law behavior. In particular, in the extreme
multi-band case, when some of the Fermi-surface sheets remain
normal (which might be the case in CePt$_3$Si \cite{exp-CePtSi}),
then $R={\rm const}+aT$ for line nodes, and $R={\rm const}+aT^2$
for point nodes. While the residual relaxation comes from the
gapless excitations on the unpaired sheets, the unusual power laws
are due to the inter-band contributions to $R$ \cite{SM05}. At
temperatures close to $T_c$, there might still exist the
Hebel-Slichter peak due to a weak singularity in
$N_\sigma(\omega)$ (for instance, for a $d$-wave gap the
singularity is logarithmic, leading to the peak of a much smaller
magnitude than in the isotropic case). In general, the presence
and the magnitude of the Hebel-Slichter peak depend on the gap
anisotropy and the band structure.

If the orbital and the spin-dipolar terms are not negligible, then
the calculations become more involved. The hyperfine field at the
nucleus can be written quite generally as a bilinear combination
of the band electron operators:
\begin{equation}
\label{H hf general}
    \bm{h}=\frac{1}{V}\sum\limits_{ij,\bk\bk'}
    \bm{\Lambda}_{ij}(\bk,\bk')c_{i,\bk}^\dagger
    c_{j,\bk'},
\end{equation}
where $\bm{\Lambda}$ is a pseudovector function,
$\bm{\Lambda}^\dagger=\bm{\Lambda}$ [for the Fermi contact
interaction considered above, it is factorized:
$\bm{\Lambda}_{ij}(\bk,\bk')=(\alpha/2)
u^{*}_{i,\bk,\sigma}\bm{\sigma}_{\sigma\sigma'}u_{j,\bk',\sigma'}$].
Under time reversal, $c^\dagger_{i,\bk}\to
t_i(\bk)c^\dagger_{i,-\bk}$ and $\bm{h}\to-\bm{h}$, which gives
the following property:
\begin{equation}
\label{Lambda symmetry}
    \bm{\Lambda}_{ij}(-\bk,-\bk')=-\bm{\Lambda}_{ji}(\bk',\bk)t_i(\bk)t^*_j(\bk').
\end{equation}
The symmetry of the Hamiltonian with respect to the point group
rotations and reflections imposes some additional constraints on
the $\bm{\Lambda}$'s, which shall not be discussed here.

Assuming that both $\bm{\Lambda}_{ij}(\bk,\bk')$ and
$\tilde\Delta_{i,\bk}$ are weakly dependent on $\xi_{i,\bk}$,
using the relations (\ref{Lambda symmetry}), and repeating all the
calculation steps from the Fermi-contact case above, we obtain
\begin{equation}
\label{R general}
    \frac{R_s}{R_n}=2\int\limits_0^\infty d\omega
    \left(-\frac{\partial f}{\partial\omega}\right)
    \frac{\sum_{ij}N_{F,i}N_{F,j}A^{(s)}_{ij}(\omega)}{\sum_{ij}
    N_{F,i}N_{F,j}A^{(n)}_{ij}},
\end{equation}
where
\begin{eqnarray}
\label{Aij}
    &&A^{(s)}_{ij}(\omega)=\Biggl\langle|\Lambda^-_{ij}(\bk,\bk')|^2\nonumber\\
    &&\quad\qquad\times\frac{\omega^2+\re[\tilde\Delta^*_{i,\bk}\tilde\Delta_{j,\bk'}]}{
    \sqrt{\omega^2-|\tilde\Delta_{i,\bk}|^2}
    \sqrt{\omega^2-|\tilde\Delta_{j,\bk'}|^2}}\Biggr\rangle_{ij},\\
    &&A^{(n)}_{ij}=\left\langle|\Lambda^-_{ij}(\bk,\bk')|^2\right\rangle_{ij}.\nonumber
\end{eqnarray}
The angular brackets here denote the average over the Fermi
surface in the $i$th and $j$th bands, and
$\Lambda^{\pm}=\Lambda^x\pm i\Lambda^y$ (recall that the spin
quantization axis is along the external field $\bm{H}$). One can
see from Eq. (\ref{Aij}) that $A^{(s)}_{ij}(\omega)\propto
\omega^2$ ($\omega^4$) at $\omega\to 0$ for line (point) nodes,
assuming non-vanishing anisotropic gaps in all bands and
neglecting the possibility of the vertex
$\Lambda^-_{ij}(\bk,\bk')$ going to zero accidentally at the gap
nodes. Therefore, the low-$T$ behavior of the relaxation rate is
characterized by the familiar exponents: $R\propto T^2$ for line
nodes, and $R\propto T^4$ for point nodes.

In conclusion, we calculated the NMR relaxation rate in a
non-centrosymmetric superconductor with strong spin-orbit
coupling, using the exact band representation of the gap functions
and the spin density operators. The temperature dependence of
$T_1^{-1}$ is determined by the interplay of intra-band and
inter-band terms, with the local densities of states weighing the
contributions from different bands. The only robust qualitative
conclusion that can be drawn from our results is that the
low-temperature behavior of $T_1^{-1}$ is determined by the same
power laws as in the centrosymmetric case: $T_1^{-1}\propto T^3$
for line nodes, and $T_1^{-1}\propto T^5$ for point nodes.

The determination of the gap symmetry in CePt$_3$Si using the NMR
relaxation data, beyond the conclusion there are likely lines of
nodes somewhere on the Fermi surface, does not seem to be possible
at the moment, since it would require a detailed knowledge of the
band structure and the local densities of states. The absence or
presence of the Hebel-Slichter peak, especially of the magnitude
observed in the experiments, cannot be used as an argument for or
against unconventional superconductivity. In addition, the
experiments were done at finite fields, in the presence of
vortices, which might obscure information about the underlying gap
symmetry and complicate the comparison with the theory presented
here.

\emph{Note added}: In a very recent theoretical investigation of
the NMR relaxation rate in CePt$_3$Si \cite{Haya05}, a different
approach was used, in which the spin-orbit band splitting is
described by the Rashba model, and the gap function has both
singlet and triplet components. In order to explain the observed
$T^3$ behavior of $T_1^{-1}$, the parameters of the model had to
be fine-tuned for the gap to have accidental lines of nodes.

\acknowledgements

The author is grateful to B. Mitrovi\'c for numerous stimulating
discussions. This work was supported by the Natural Sciences and
Engineering Research Council (NSERC) of Canada.

\end{document}